\documentclass[showpacs,twocolumn,prl]{revtex4}
\usepackage{amsfonts}
\usepackage{amssymb}
\usepackage{amsmath}
\usepackage{graphicx}
\usepackage{epsfig}

\begin{document}

\title{Thermal States as Universal Resources for Quantum Computation with
Always-on Interactions}
\author{Ying Li$^{1}$}
\author{Daniel E. Browne$^{1,2}$}
\author{Leong Chuan Kwek$^{1,3}$}
\author{Robert Raussendorf$^{4}$}
\author{Tzu-Chieh Wei$^{4}$}
\affiliation{$^{1}$Centre for Quantum Technologies, National University of Singapore, 2
Science Drive 3, Singapore}
\affiliation{$^{2}$Department of Physics and Astronomy, University College London, Gower
Street, London WC1E 6BT, United Kingdom}
\affiliation{$^{3}$National Institute of Education and Institute of Advanced Studies,
Nanyang Technological University, 1 Nanyang Walk, Singapore}
\affiliation{$^{4}$Department of Physics and Astronomy, University of British Columbia,
Vancouver, British Columbia V6T 1Z1, Canada}
\date{\today}

\begin{abstract}
Measurement-based quantum computation utilizes an initial entangled resource
state and proceeds with subsequent single-qubit measurements. It is
implicitly assumed that the interactions between qubits can be switched off
so that the dynamics of the measured qubits do not affect the computation.
By proposing a model spin Hamiltonian, we demonstrate that measurement-based
quantum computation can be achieved on a thermal state with always-on
interactions. Moreover, computational errors induced by thermal fluctuations
can be corrected and thus the computation can be executed fault-tolerantly
if the temperature is below a threshold value.
\end{abstract}

\maketitle

\textit{Introduction.- }A quantum computer can solve certain problems
considered hard for a classical computer with an exponential speedup~\cite%
{NielsenChuang00}. Standard quantum computing uses unitary evolution as a
basic mechanism for information processing. Another paradigm is
measurement-based quantum computing (MBQC), in which one processes quantum
information by single-particle operations and measurements only, on a
nontrivial entangled state~\cite{Oneway}. Such entangled states serve as
universal resources of MBQC~\cite{Universal}. The first identified universal
resource was the cluster state. It can be obtained as the unique ground
state of a Hamiltonian with five-body interactions \cite{5-body}, but can
never occur as the unique ground state of any two-body Hamiltonian~\cite%
{Nielsen}. Fortunately, there exist universal resources that are the unique
ground states of two-body Hamiltonians, albeit with particles of local
Hilbert space larger than that of a qubit. These two-body Hamiltonians
include the tricluster model~\cite{6level}, an Affleck-Kennedy-Lieb-Tasaki
(AKLT)-like model~\cite{4level}, the two dimensional AKLT model~\cite%
{AKLTstate,2DAKLT} and a quadratic Hamiltonian of continuous variables~\cite%
{CV}. However, in order to use the ground state of a system as a universal
resource, one usually needs to switch off interactions of the system
sequentially~\cite{6level,2DAKLT,4level,1DAKLT}. Otherwise, the desirable
quantum correlations could be destroyed due to the time evolution of the
state via interactions. Therefore, in previous proposals, MBQC based on
ground states requires not only single-particle operations and measurements
but also a good control of interactions. In this paper, we find that it is
possible to remove this extra requirement, i.e., MBQC can be performed with
always-on interactions.

%%%%%%%%%%%%%%%%%%%%%%%%%%%%%%%%%%%%%%%%%%%%%%%%%%%%%%%%%%%%%%%%%%%%%%%%%%%%%%%

\begin{figure}[tbp]
\includegraphics[bb=50 445 550 815, width=7.5 cm]{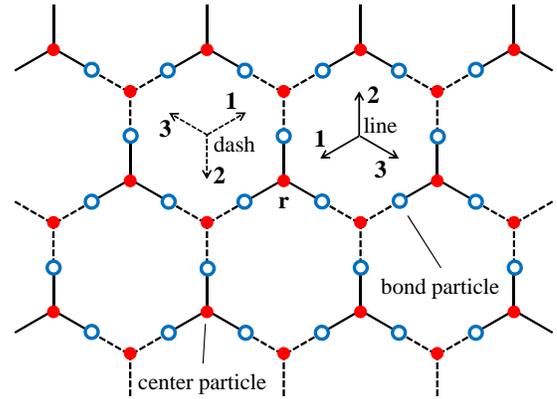}
\caption{The two-dimensional system composed of spin-$3/2$ particles. The
system is a hexagonal lattice, where center particles (red round) are
located on vertices, while bond particles (blue ring) are located on edges.
There are two kinds of interactions between center particles and bond
particles, $V_{\mathrm{line}}$ (line) and $V_{\mathrm{dash}}$ (dash line). $%
\mathbf{r}$ denotes the position of a center particle, and the vectors
between the center particle and its three interacting bond particles are $%
\mathbf{1},\mathbf{2},\mathbf{3}$ respectively.}
\label{system}
\end{figure}

%%%%%%%%%%%%%%%%%%%%%%%%%%%%%%%%%%%%%%%%%%%%%%%%%%%%%%%%%%%%%%%%%%%%%%%%%%%%%%%

To this end, we propose a two-dimensional (2D) system and a
three-dimensional (3D) system, whose ground states are universal resources
for MBQC. We show that 2D and 3D systems can be generalized to a family of
similar models. These spin models may be realized in physical systems such
as cold atoms~\cite{coldatom}, polar molecules~\cite{molecules}, trapped
ions~\cite{ions} and Josephson junction array~\cite{junction}. We construct
a ground state as a universal resource for MBQC by showing that the ground
state can be converted into a cluster state by single-particle operations
and measurements \cite{2DAKLT,Reduction}. In practice, one obtains a thermal
state instead of the ground state as a universal resource for MBQC. Thus an
energy gap is needed to protect the state from thermal fluctuations, which
is indeed the case in our model. However, it is not clear how high a
temperature can be tolerated before the state would no longer be a universal
resource of MBQC. Therefore, we investigate their thermal states and find
that computational errors in MBQC induced by thermal fluctuations can be
corrected as long as the temperature is below a certain value.

\textit{2D System.- }The 2D system is shown in Fig. \ref{system}, which is a
hexagonal lattice with one more particle on each edge. The system is
composed of spin-$3/2$ particles, in which particles on edges are called
bond particles, while others are called center particles. Particles are
combined by two types of interactions
\begin{eqnarray}
V_{\mathrm{line}} &=&\Delta (S_{\mathbf{c}}^{x}A_{\mathbf{b}}^{x}+S_{\mathbf{%
c}}^{y}A_{\mathbf{b}}^{y}+S_{\mathbf{c}}^{z}A_{\mathbf{b}}^{z}),
\label{line} \\
V_{\mathrm{dash}} &=&\Delta (S_{\mathbf{c}}^{x}B_{\mathbf{b}}^{x}+S_{\mathbf{%
c}}^{y}B_{\mathbf{b}}^{y}+S_{\mathbf{c}}^{z}B_{\mathbf{b}}^{z}),
\label{dash}
\end{eqnarray}%
where $S_{\mathbf{c}}^{\alpha }$ is the spin operator of the corresponding
center particle, and $A_{\mathbf{b}}^{\alpha },B_{\mathbf{b}}^{\alpha }$ are
operators of the corresponding bond particle \cite{operators}. Operators of
bond particles satisfy commutation relations $[I_{\mathbf{b}}^{\alpha },J_{%
\mathbf{b}}^{\beta }]=i\delta _{IJ}\epsilon _{\alpha \beta \gamma }I_{%
\mathbf{b}}^{\gamma }$ and $\overrightarrow{I}_{\mathbf{b}}^{2}=3/4$, where $%
I,J=A,B$, and $\alpha ,\beta ,\gamma =x,y,z$. Therefore, $A_{\mathbf{b}%
}^{\alpha }$ and $B_{\mathbf{b}}^{\alpha }$ are two sets of independent spin-%
$1/2$ operators.

The Hamiltonian of the system is $H=\sum_{\mathbf{r}}h_{\mathbf{r}}$, where $%
h_{\mathbf{r}}=\Delta \overrightarrow{S}_{\mathbf{r}}\cdot \overrightarrow{I}%
_{\mathbf{r}}$, $\overrightarrow{I}_{\mathbf{r}}=\overrightarrow{I}_{\mathbf{%
r}+\mathbf{1}}+\overrightarrow{I}_{\mathbf{r}+\mathbf{2}}+\overrightarrow{I}%
_{\mathbf{r}+\mathbf{3}}$, and $\mathbf{r}$ denotes the position of a center
particle, $\mathbf{r}+\mathbf{a}$ denotes the position of one bond particle
interacting with the center particle $\mathbf{r}$. Here, $\{\mathbf{a}\}$
depends on $I\in \{A,B\}$ as shown in Fig. \ref{system}.

\textit{Ground state and energy gap.- }We can rewrite $h_{\mathbf{r}}$ as $%
h_{\mathbf{r}}=\Delta (\overrightarrow{T}_{\mathbf{r}}^{2}-\overrightarrow{S}%
_{\mathbf{r}}^{2}-\overrightarrow{I}_{\mathbf{r}}^{2})/2$, where $%
\overrightarrow{T}_{\mathbf{r}}=\overrightarrow{S}_{\mathbf{r}}+%
\overrightarrow{I}_{\mathbf{r}}$. Here, $\overrightarrow{S}_{\mathbf{r}}$, $%
\overrightarrow{I}_{\mathbf{r}}$ and $\overrightarrow{T}_{\mathbf{r}}$ all
satisfy commutation relations of spin operators. Therefore,
\begin{equation}
h_{\mathbf{r}}=\frac{\Delta }{2}[T_{\mathbf{r}}(T_{\mathbf{r}}+1)-S_{\mathbf{%
r}}(S_{\mathbf{r}}+1)-I_{\mathbf{r}}(I_{\mathbf{r}}+1)],  \label{hr}
\end{equation}%
where $S_{\mathbf{r}}=3/2$ and $I_{\mathbf{r}}=1/2$ or $3/2$. When $I_{%
\mathbf{r}}=1/2$, $T_{\mathbf{r}}=1,2$. When $I_{\mathbf{r}}=3/2$, $T_{%
\mathbf{r}}=0,1,2,3$. One can get the minimum energy by taking $I_{\mathbf{r}%
}=3/2$ and $T_{\mathbf{r}}=0$, which means the ground state, $\left\vert
\mathrm{g}\right\rangle _{\mathbf{r}}$, of $h_{\mathbf{r}}$ has a total spin
$0$. The energy difference between the ground state and the first excited
state is $\Delta $. Because these $h_{\mathbf{r}}$ are independent with each
other, the ground state of the whole system is $\left\vert \mathrm{G}%
\right\rangle =\bigotimes_{\mathbf{r}}\left\vert \mathrm{g}\right\rangle _{%
\mathbf{r}}$ and protected by an energy gap $\Delta $. The energy gap only
depends on the interaction constant, and does not vanish in the
thermodynamic limit.

\textit{POVM and GHZ state - }As the first step of MBQC on the ground state,
the POVM $\mathbf{I}=\sum_{\alpha =x,y,z}F^{\alpha \dagger }F^{\alpha }$ is
performed on center particles. Here, $F^{\alpha }=(S_{\mathbf{r}}^{\alpha
2}-1/4)/\sqrt{6}$, which projects the center spin into the subspace spanned
by two states with maximum spin component in the $\alpha $ direction.
Because the ground state $\left\vert \mathrm{g}\right\rangle _{\mathbf{r}}$
has a total spin $0$, all three spin-$I_{\mathbf{r}+\mathbf{a}}$ are
antiparallel with the center spin-$S_{\mathbf{r}}$. Therefore, the POVM
projects the state $\left\vert \mathrm{g}\right\rangle _{\mathbf{r}}$ into a
GHZ state, e.g., for the outcome $z$, the output state is $\left\vert
\mathrm{ghz}\right\rangle _{\mathbf{r}}=(|\widetilde{0}000\rangle +|%
\widetilde{1}111\rangle )/\sqrt{2}$, where $|\widetilde{0}\rangle =-|S_{%
\mathbf{r}}^{z}=3/2\rangle $, $|\widetilde{1}\rangle =|S_{\mathbf{r}%
}^{z}=-3/2\rangle $ are the state of the center spin, and $\left\vert
0\right\rangle $ ($\left\vert 1\right\rangle $) is the eigenstate of $I_{%
\mathbf{r}+\mathbf{a}}^{z}$ with eigenvalue $-1/2$ ($1/2$). The state $%
\left\vert \mathrm{g}\right\rangle _{\mathbf{r}}$ is an isotropic state.
Therefore, all outcomes are equivalent to the outcome $z$ up to a set of
single-particle operations $U(\widehat{\alpha })=\exp [i\overrightarrow{T}%
\cdot \overrightarrow{n}(\widehat{\alpha })]$, where $\alpha $ is the
outcome and $\overrightarrow{n}(\widehat{\alpha })=\widehat{\alpha }\times
\widehat{z}\arcsin (|\widehat{\alpha }\times \widehat{z}|)/|\widehat{\alpha }%
\times \widehat{z}|$. Then, the state of the whole system after POVMs and
single-particle operations is $\left\vert \mathrm{\{ghz\}}\right\rangle
=\bigotimes_{\mathbf{r}}\left\vert \mathrm{ghz}\right\rangle _{\mathbf{r}}$,
which can also be described by a set of stabilizers, $W_{\mathbf{r}}=X_{%
\mathbf{r}}\prod_{\mathbf{a}=\mathbf{1},\mathbf{2},\mathbf{3}}2I_{\mathbf{r}+%
\mathbf{a}}^{x}$ and $W_{\mathbf{r},\mathbf{r}+\mathbf{a}}=2Z_{\mathbf{r}}I_{%
\mathbf{r}+\mathbf{a}}^{z}$ for all $\mathbf{r}$ and $\mathbf{a}$.\ $%
\left\vert \mathrm{\{ghz\}}\right\rangle $ is the eigenstate with eigenvalue
$1$ of all of these stabilizers. Here, $X$, $Y$ and $Z$ are Pauli operators
of the qubit $\{|\widetilde{0}\rangle ,|\widetilde{1}\rangle \}$.

%%%%%%%%%%%%%%%%%%%%%%%%%%%%%%%%%%%%%%%%%%%%%%%%%%%%%%%%%%%%%%%%%%%%%%%%%%%%%%%

\begin{figure}[tbp]
\includegraphics[bb=70 280 530 675, width=7.5 cm]{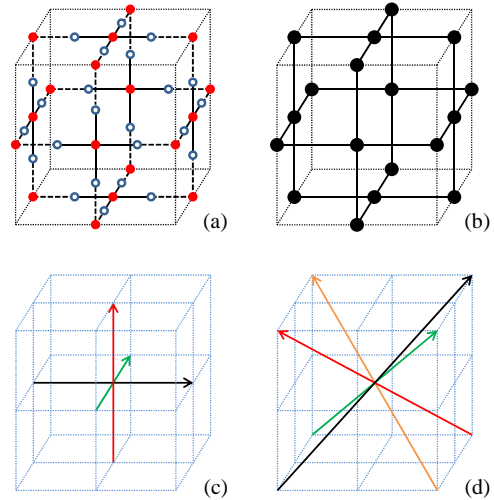}
\caption{The three-dimensional system. (a) The elementary cell of the
system. The system is composed of spin-$2$ particles and spin-$3/2$
particles. Spin-$2$ particles are center particles (red round), and spin-$%
3/2 $ particles are bond particles (blue ring). (b) The elementary cell of
the three dimensional topology-protected cluster state, which can be
prepared by single-particle operations and measurements on the ground state
of the 3D system. (c) Three directions for $k\leq 3$ of the POVM on spin-$2$
particles, which are orthogonal with each other and passing face centers of
a cube. (d) Four directions for $k\geq 4$ of the POVM on spin-$2$ particles,
which are along body diagonals of the same cube.}
\label{3D}
\end{figure}

%%%%%%%%%%%%%%%%%%%%%%%%%%%%%%%%%%%%%%%%%%%%%%%%%%%%%%%%%%%%%%%%%%%%%%%%%%%%%%%

\textit{Cluster state and universality of the ground state - }By measuring
physical quantities $A_{\mathbf{b}}^{x}B_{\mathbf{b}}^{z}$ and $A_{\mathbf{b}%
}^{z}B_{\mathbf{b}}^{x}$ on bond particles, the state $\left\vert \mathrm{%
\{ghz\}}\right\rangle $ can be projected \cite{GottesmanChuang09,fusion} (or
\textquotedblleft fused\textquotedblright ) into a hexagonal cluster state,
which has the same lattice with center particles. Eigenstates of $A_{\mathbf{%
b}}^{x}B_{\mathbf{b}}^{z}$ and $A_{\mathbf{b}}^{z}B_{\mathbf{b}}^{x}$, which
are measurement basis, can be found in Ref. \cite{operators}. Considering a
product of stabilizers
\begin{equation}
W_{\mathbf{r}}\prod_{\mathbf{a}=\mathbf{1},\mathbf{2},\mathbf{3}}W_{\mathbf{r%
}+2\mathbf{a},\mathbf{r}+\mathbf{a}}=X_{\mathbf{r}}\prod_{\mathbf{a}=\mathbf{%
1},\mathbf{2},\mathbf{3}}Z_{\mathbf{r}+2\mathbf{a}}(4A_{\mathbf{r}+\mathbf{a}%
}^{x}B_{\mathbf{r}+\mathbf{a}}^{z}),  \label{stabilizer}
\end{equation}%
one can get a new stabilizer by replacing $A_{\mathbf{b}}^{x}B_{\mathbf{b}%
}^{z}$ with outcomes. In Eq. (\ref{stabilizer}), we have taken the case $I_{%
\mathbf{r}+\mathbf{a}}^{x}=A_{\mathbf{r}+\mathbf{a}}^{x}$ as an example, and
the result is the same for $I_{\mathbf{r}+\mathbf{a}}^{x}=B_{\mathbf{r}+%
\mathbf{a}}^{x}$. The new stabilizers define a hexagonal cluster state\ on
center particles up to a Pauli frame, which can be corrected by
single-particle operations \cite{Oneway}. The hexagonal cluster state is a
universal resource for MBQC~\cite{Universal}. Then, universal MBQC can be
performed on center particles.

\textit{3D system and topology-protected cluster state - }Following the idea
of 2D system, we propose a 3D system, whose ground state is also a universal
resource for MBQC. The system is shown in Fig. \ref{3D} (a), which is
composed by spin-$2$ particles and spin-$3/2$ particles, where center
particles are spin-$2$ particles and bond particles are spin-$3/2$
particles. The interactions between particles are the same as Eq. (\ref{line}%
) and (\ref{dash}). Therefore, the Hamiltonian of the 3D system has the same
form as 2D system, $H=\sum_{\mathbf{r}}h_{\mathbf{r}}$, $h_{\mathbf{r}%
}=\Delta \overrightarrow{S}_{\mathbf{r}}\cdot \overrightarrow{I}_{\mathbf{r}%
} $, where $\overrightarrow{I}_{\mathbf{r}}=\overrightarrow{I}_{\mathbf{r}+%
\mathbf{1}}+\overrightarrow{I}_{\mathbf{r}+\mathbf{2}}+\overrightarrow{I}_{%
\mathbf{r}+\mathbf{3}}+\overrightarrow{I}_{\mathbf{r}+\mathbf{4}}$. Here, $\{%
\mathbf{r}+\mathbf{a}\}$ denote four bond particles interacting with the
center particle $\mathbf{r}$.

In the 3D system, one can get the minimum energy of $h_{\mathbf{r}}$ by
taking $I_{\mathbf{r}}=2$ and $T_{\mathbf{r}}=0$ in Eq. (\ref{hr}).
Therefore, in the 3D system, the ground state of each $h_{\mathbf{r}}$ is an
isotropic state with a total spin $0$. The energy difference between the
ground state and the first excited state is $\Delta $, which means the 3D
system is also gapped.

The ground state of the 3D system can be reduced to a 3D cluster state, as
shown in Fig. \ref{3D} (b). Firstly, center particles are measured as $%
\mathbf{I}=\sum_{k=1}^{7}F^{\dag }(\widehat{\alpha }_{k})F(\widehat{\alpha }%
_{k})$, which is a POVM with seven outcomes. Here, $F(\widehat{\alpha }_{k})=%
\sqrt{N_{k}}P(\widehat{\alpha }_{k})$, and
\begin{equation}
P(\widehat{\alpha })=\left\vert \widehat{\alpha };2\right\rangle
\left\langle \widehat{\alpha };2\right\vert +\left\vert \widehat{\alpha }%
;-2\right\rangle \left\langle \widehat{\alpha };-2\right\vert
\label{projection}
\end{equation}%
projects the center spin into the subspace spanned by two states with
maximum spin component in the $\widehat{\alpha }$ direction. $\left\vert
\widehat{\alpha };m\right\rangle $ is the eigenstate of $\widehat{\alpha }%
\cdot \overrightarrow{S}_{\mathbf{c}}$ with eigenvalue $m$. $N_{k}=1/3$ for $%
k\leq 3$ and $N_{k}=3/8$ for $k\geq 4$. The seven directions are shown in
Fig. \ref{3D} (c) and (d). Because four spins $\{I_{\mathbf{r}+\mathbf{a}}\}$
are all antiparallel with the center spin-$S_{\mathbf{r}}$, the output
states of the POVM are GHZ states. These GHZ states are equivalent to the
GHZ state of outcome $z$, up to single-particle operations $U(\widehat{%
\alpha })$. Therefore, POVMs on center particles, with $U(\widehat{\alpha })$
together, can transform the ground state to a state stabilized by $W_{%
\mathbf{r}}=X_{\mathbf{r}}\prod_{\mathbf{a}}2I_{\mathbf{r}+\mathbf{a}}^{x}$
and $W_{\mathbf{r},\mathbf{r}+\mathbf{a}}=2Z_{\mathbf{r}}I_{\mathbf{r}+%
\mathbf{a}}^{z}$, where $\mathbf{a}=\mathbf{1},\mathbf{2},\mathbf{3},\mathbf{%
4}$.

Measuring physical quantities $A_{\mathbf{b}}^{x}B_{\mathbf{b}}^{z}$ and $A_{%
\mathbf{b}}^{z}B_{\mathbf{b}}^{x}$, one can generate a new set of
stabilizers $X_{\mathbf{r}}\prod_{\mathbf{a}}Z_{\mathbf{r}+2\mathbf{a}}$,
which defines a 3D cluster state on center particles, as shown in Fig. \ref%
{3D} (b). On the 3D cluster state, quantum correlations are protected
topologically and fault tolerant quantum computing can be simulated using
topological error correction \cite{R.Raussendorf}.

%%%%%%%%%%%%%%%%%%%%%%%%%%%%%%%%%%%%%%%%%%%%%%%%%%%%%%%%%%%%%%%%%%%%%%%%%%%%%%%

\begin{figure}[tbp]
\includegraphics[bb=30 200 555 640, width=7.5 cm]{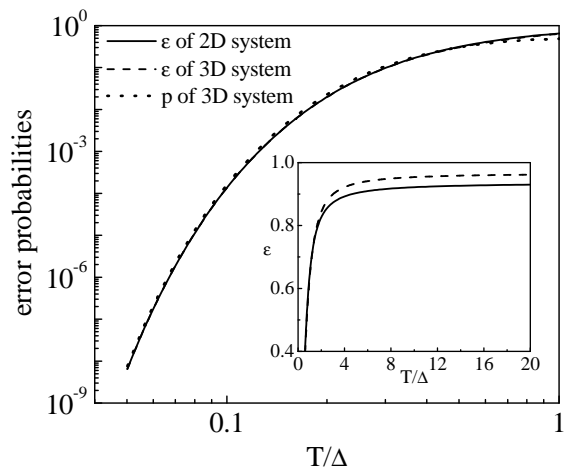}
\caption{Error probabilities $\protect\epsilon $ on a GHZ state and $p$ on
the cluster state as functions of the temperature. For 2D system the GHZ
state is a four-qubit state, and for 3D system the GHZ state is a five-qubit
state. $\protect\epsilon $ are almost the same for 2D and 3D systems when
the temperature $T<\Delta $. When $T/\Delta =0.2$, $p=3\%$, which means
errors induced by a lower temperature are tolerable by using the topological
error correction algorithm on the 3D system.}
\label{TE}
\end{figure}

%%%%%%%%%%%%%%%%%%%%%%%%%%%%%%%%%%%%%%%%%%%%%%%%%%%%%%%%%%%%%%%%%%%%%%%%%%%%%%%

\textit{Thermal computational errors and error correction - }We have proved
the ground states of 2D and 3D systems are universal resources for MBQC.
However, in practice, a system cannot reach the exact ground state, but
rather a thermal state at finite temperature. Thermal fluctuations can
reduce the quantum correlations on ground states and induce computational
errors on the cluster state, which will be used for MBQC. The thermal state
is the Gibbs state $\rho =Z^{-1}e^{-\beta H}$, where $Z=\mathrm{tr}e^{-\beta
H}$, $\beta =1/T$ is the temperature, and $\rho $ can be rewritten as $\rho
=\prod_{\mathbf{r}}\rho _{\mathbf{r}}$. Here $\rho _{\mathbf{r}}=Z_{\mathbf{r%
}}^{-1}e^{-\beta h_{\mathbf{r}}}$ is the Gibbs state of $h_{\mathbf{r}}$.
After the POVM and $U(\widehat{\alpha })$, the state $\rho _{\mathbf{r}}$ is
transformed into $\sigma _{\mathbf{r}}=F\rho _{\mathbf{r}}F^{\dag }/\mathrm{%
tr}(F\rho _{\mathbf{r}}F^{\dag })$, where $F=F^{z}$ for the 2D system and $%
F=F(\widehat{z})$ for the 3D system. At an absolute zero temperature, $%
\sigma _{\mathbf{r}}=\left\vert \mathrm{ghz}\right\rangle \left\langle
\mathrm{ghz}\right\vert _{\mathbf{r}}$ is the desired GHZ state. Here, $%
\left\vert \mathrm{ghz}\right\rangle _{\mathbf{r}}$ is a GHZ state of four
qubits for the 2D system and five qubits for the 3D system.

The post-POVM state $\sigma _{\mathbf{r}}$ at finite $T$ is only
approximately a GHZ state, i.e., is equivalent to a perfect GHZ state
affected by errors. The probability of an error occurring on the post-POVM
state is $\epsilon =1-\mathcal{F}$, where $\mathcal{F}=\mathrm{tr}(\sigma _{%
\mathbf{r}}\left\vert \mathrm{ghz}\right\rangle \left\langle \mathrm{ghz}%
\right\vert _{\mathbf{r}})$ is the fidelity of the GHZ state, as shown in
Fig. \ref{TE}. Those errors are propagated under the measurements of the
bond particles and subsequent correction operations \cite{YingLi2010}. The
resulting error superoperators act on the yet unmeasured center particles,
and have the following properties: (1) there is one independent error
superoperator ${\mathcal{E}}_{\mathbf{r}}$ for every $\mathbf{r}$, (2) ${%
\mathcal{E}}_{\mathbf{r}}$ acts at the locations $\mathbf{r}$ and $\{\mathbf{%
r}+2\mathbf{a},\forall \mathbf{a}\}$. Where the center particles are
measured in the $X$-basis for the purpose of topological error correction on
the 3D cluster state (in most of the cluster), there arise two further
simplifications: (3) All errors are equivalent to $Z$-errors or the
identity, and (4) Correlations between errors on neighboring center
particles can be discarded. The latter arises because errors at $\mathbf{r}$
and at $\mathbf{r}+2\mathbf{a}$ are corrected by different error-correction
procedures running independently of another \cite{R.Raussendorf}.

On the 3D cluster state, for each $\mathbf{r}$, the resulting error is ${%
\mathcal{E}}_{\mathbf{r}}=E_{1}\circ E_{2}$, with $E_{1}=(1-p_{1})+p_{1}[Z_{%
\mathbf{r}}]$ and $E_{2}=(1-p_{2}-p_{3})+p_{2}/4\sum_{\mathbf{a}}[Z_{\mathbf{%
r}+2\mathbf{a}}]+p_{3}/6\sum_{\mathbf{a},\mathbf{a}^{\prime }}[Z_{\mathbf{r}%
+2\mathbf{a}}Z_{\mathbf{r}+2\mathbf{a}^{\prime }}]$. Therein, the error
probabilities $p_{1}$, $p_{2}$ and $p_{3}$ depend on the temperature $T$. If
$p_{3}\ll p_{1},p_{2}\ll 1$, then the local errors are almost independent
and the error level is described by an effective local error probability $%
p\simeq p_{1}+p_{2}+2p_{3}$. Error-correction is possible if $p<3\%$ \cite%
{threshold}, which translates into a threshold temperature $T_{t}=0.2\Delta $%
; See Fig. \ref{TE}. At that point, $p_{1},p_{2}\sim 10^{-2}$ and $%
p_{3}=10^{-6}$, justifying the assumption of uncorrelated local errors.

\textit{MBQC with always-on interactions - }In practical application, one
can convert the initial state, usually a thermal state, to a cluster state
one qubit at a time. Once we need the qubit $\mathbf{r}$, we can apply POVMs
on the center particle $\mathbf{r}$ and its neighboring center particles $\{%
\mathbf{r}+2\mathbf{a}\}$. Based on outcomes of POVMs, single-particle
operations $U(\widehat{\alpha })$ are chosen. Then, bond particles $\{%
\mathbf{r}+\mathbf{a}\}$ are measured, and outcomes are used to correct the
Pauli frame of qubit $\mathbf{r}$. No further operation is needed on any
other particle in order to convert the center particle $\mathbf{r}$ to a
qubit on the cluster state.

With always-on interactions, we need to consider the time evolution driven
by the time-independent Hamiltonian. Since the initial state is not
converted to the cluster state simultaneously, there are some untouched
particles. They remain in the initial thermal state, which is close to the
ground state due to the existence of the energy gap. Other particles evolve
with the Hamiltonian, and their quantum correlations will be changed under
time evolution. Fortunately, the time evolution is periodic with a period $%
4\pi /\Delta $ for the 2D system and $2\pi /\Delta $ for the 3D system.\
Then, one can perform operations on these particles at the revival time of
quantum correlations, $t=4n\pi /\Delta $ and $t=2n\pi /\Delta $ for 2D and
3D systems respectively, where $t=0$ is the time of the first operation on
the particle and its interaction particles, and $n=0,1,2,\ldots $. If we
assume only one operation can be performed on each particle at one revival
time, particles can be measured out before $n=6$. Therefore, the MBQC can be
performed on our proposed systems with always-on interactions. Here,
operations are required to occur precisely. We remark that errors due to
timing imprecision can also be analyzed similarly to thermal errors.

\textit{Discussion - }In summary, we proposed a 2D and a 3D gapped system,
whose ground state is entangled based on a factorized Hamiltonian. With a
factorized Hamiltonian, quantum computing can be performed without the need
to switch off interactions. The ground state can be reduced to a
deterministic cluster state, in contrast to AKLT like systems where cluster
states are obtained with stochastic structures. Errors induced by thermal
fluctuations can be corrected as long as the temperature is below a critical
threshold. There are other choices of $A_{\mathbf{b}}^{\alpha }$ and $B_{%
\mathbf{b}}^{\alpha }$ that satisfy conditions of spin-$1/2$ operators. By
replacing the center particle with different spin systems, i.e. spin-$m/2$
particles, one can get different spatial connectivities that each qubit is
connected to $m$ other qubits in the cluster state. Thereby, it can be
generalized to 3D and more complicated configurations.

%\begin{acknowledgements}

Y. L. thanks Simon C. Benjamin for helpful discussions. This work was
supported by National Research Foundation \& Ministry of Education,
Singapore (L.-C. K., D. E. B. and Y. L.), Leverhulme Trust (D. E. B), NSERC
of Canada, MITACS (R. R. and T.-C. W.), Cifar and Sloan Foundation (R. R.).

%\end{acknowledgements}


\begin{thebibliography}{99}
\bibitem{operators} $A_{\mathbf{b}}^{x}=(S_{\mathbf{b}}^{-2}+S_{\mathbf{b}%
}^{+2})/4\sqrt{3}$, $A_{\mathbf{b}}^{y}=i(S_{\mathbf{b}}^{-2}-S_{\mathbf{b}%
}^{+2})/4\sqrt{3}$, $A_{\mathbf{b}}^{z}=S_{\mathbf{b}}^{z}(13-4S_{\mathbf{b}%
}^{z2})/12$ and $B_{\mathbf{b}}^{x}=(S_{\mathbf{b}}^{x}+4S_{\mathbf{b}%
}^{z}S_{\mathbf{b}}^{x}S_{\mathbf{b}}^{z})/4\sqrt{3}$, $B_{\mathbf{b}%
}^{y}=(S_{\mathbf{b}}^{y}+4S_{\mathbf{b}}^{z}S_{\mathbf{b}}^{y}S_{\mathbf{b}%
}^{z})/4\sqrt{3}$, $B_{\mathbf{b}}^{z}=S_{\mathbf{b}}^{z}(4S_{\mathbf{b}%
}^{z2}-7)/6$. The eigenstate with $A_{\mathbf{b}}^{x}B_{\mathbf{b}%
}^{z}=(1-2\mu )/4$ and $A_{\mathbf{b}}^{z}B_{\mathbf{b}}^{x}=(1-2\nu )/4$ is
$\left\vert \mu ,\nu \right\rangle =(\left\vert 3/2\right\rangle +i^{2\nu
}\left\vert 1/2\right\rangle +i^{2\mu }\left\vert -1/2\right\rangle -i^{2\mu
}i^{2\nu }\left\vert -3/2\right\rangle )/2$, where $\mu ,\nu =0,1$ and $S_{%
\mathbf{b}}^{z}\left\vert m\right\rangle =m\left\vert m\right\rangle $.
\end{thebibliography}
\end{document}